\documentclass[twocolumn,showpacs,preprintnumbers,amsmath,amssymb,superscriptaddress,floatfix]{revtex4-1}
\usepackage{graphicx}% Include figure files
\usepackage{dcolumn}% Align table columns on decimal point
\usepackage{bm}% bold math

\begin{document}

%\preprint{cond-mat/???????}

\title{Resonant soft x-ray scattering from La$_{1-x}$Sr$_x$MnO$_3$ quantum wire arrays}%

\author{X. M. Chen}
  \affiliation{Department of Physics and Frederick Seitz Materials Research Laboratory, University of Illinois, Urbana, IL, 61801}
\author{E. M. Spanton}
  \affiliation{Department of Physics, Stanford University,Stanford, CA, 94305}
\author{Shuai Wang}
  \affiliation{Department of Physics and Frederick Seitz Materials Research Laboratory, University of Illinois, Urbana, IL, 61801}
\author{J. C. T. Lee}
  \affiliation{Department of Physics and Frederick Seitz Materials Research Laboratory, University of Illinois, Urbana, IL, 61801}
\author{S. Smadici}
  \affiliation{Department of Physics and Frederick Seitz Materials Research Laboratory, University of Illinois, Urbana, IL, 61801}
\author{X. Zhai}
  \affiliation{Department of Physics and Frederick Seitz Materials Research Laboratory, University of Illinois, Urbana, IL, 61801}
\author{T. Naibert}
  \affiliation{Department of Physics and Frederick Seitz Materials Research Laboratory, University of Illinois, Urbana, IL, 61801}
\author{J. N. Eckstein}
  \affiliation{Department of Physics and Frederick Seitz Materials Research Laboratory, University of Illinois, Urbana, IL, 61801}
\author{A. Bhattacharya}
  \affiliation{Center for Nanoscale Materials, Argonne National Laboratory, Argonne, IL, 60439}
\author{T. Santos}
  \affiliation{Center for Nanoscale Materials, Argonne National Laboratory, Argonne, IL, 60439}\
\author{R. Budakian}
  \affiliation{Department of Physics and Frederick Seitz Materials Research Laboratory, University of Illinois, Urbana, IL, 61801}
\author{P. Abbamonte}
  \affiliation{Department of Physics and Frederick Seitz Materials Research Laboratory, University of Illinois, Urbana, IL, 61801}
  \email{abbamonte@mrl.uiuc.edu}

\date{\today}

\begin{abstract}
We describe a strategy for using resonant soft x-ray scattering (RSXS) to study the electronic structure of transition metal oxide quantum wires.  Using electron beam lithography and ion milling, we have produced periodic, patterned arrays of colossal magnetoresistance (CMR) phase La$_{1-x}$Sr$_x$MnO$_3$ consisting of $\sim$ 5000 wires, each of which is 80 nm in width.  The scattered intensity exhibits a series of peaks that can be interpreted as Bragg reflections from the periodic structure or, equivalently, diffraction orders from the grating-like structure.  RSXS measurements at the Mn $L_3$ edge, which has a large magnetic cross section, show clear evidence for a magnetic superstructure with a commensurate period of five wires, which we interpret as commensurately modulated antiferromagnetism.  This superstructure, which is accompanied by non-trivial reorganization of the magnetization within each wire, likely results from classical dipole interactions among the wires.  We introduce a simple, exactly soluble, analytic model of the scattering that captures, semi-quantitatively, the primary features in the RSXS data; this model will act as a foundation for forthcoming, detailed studies of the magnetic structure in these systems.  

\end{abstract}

%\pacs{???,???}% PACS, the Physics and Astronomy
\maketitle

The ground state of many strongly correlated transition metal compounds is determined by a subtle competition among multiple effects and interactions, resulting in an array of nearly degenerate, competing phases with different properties.  The resulting ground state, when homogeneous, is often unstable with respect to the formation of phases with broken symmetry of either a conventional or topological nature.

A promising new approach to studying such instabilities is that of manipulating boundary conditions.  By studying the properties of materials with engineered geometry or topology, such as nanoscale quantum wires or quantum loops, new insight into the organization of electronic degrees of freedom can be obtained.  For example, it was argued by Chernyshev \cite{chernyshev} that a nanopatterned doped Mott insulator should exhibit boundary states that involve pinned stripes -- an effect that may partly explain anomalous transport phenomena observed in quantum wires of the manganites (La$_{1-x}$Pr$_x$)$_{1-y}$Ca$_y$MnO$_3$,\cite{ward2008} La$_{1-x}$Ba$_x$MnO3 \cite{nakajima}, and Pr$_x$(Ca$_{1-y}$Sr$_y$)$_{1-x}$MnO$_3$ \cite{wu}, which exhibit colossal magnetoresistance (CMR), as well as underdoped but superconducting YBa$_2$Cu$_3$O$_{7-\delta}$.\cite{bonetti}  Further, it has recently been shown that nanoscale loops of triplet superconducting Sr$_2$RuO$_4$ can stabilize the so-called half quantum vortex (HQV) state,\cite{raffiHQV} whose boundary is believed to bind a Majorana Fermion, which is a zero-energy quasiparticle that is its own anti-particle and exhibits nonabelian, anyonic statistics.  

To obtain deeper insight into such patterned, strongly correlated systems, spatially-resolved studies of their electronic structure, for example that distinguish between bulk and boundary states, are needed.  A highly sensitive and selective probe of the spatial organization of electronic states is resonant soft x-ray scattering (RSXS), which has been used in the past to study valence band order in strongly correlated materials in which translational symmetry is broken either spontaneously \cite{lsco2002, ladders2004, lbco2005, wilkins2011, scagnoli2010} or artificially via layer-by-layer synthesis.\cite{smadici2007,smadici2009,benckiser2011}  However, as a momentum space probe, RSXS is best suited to systems that are spatially periodic.  Application to structures that are aperiodic or finite in extent is not straight-forward, and requires coherent beam techniques in conjunction with a solution to the phase problem, which is an approach that is itself still a current topic of research.\cite{ikr2009,mcnulty2011,vdlaan2011}

Here we demonstrate a practical, intermediate approach, which is to study periodic arrays consisting of many identical copies of the finite structure of interest.  Unlike the scattering from a single, finite structure, which is distributed everywhere in momentum space, scattering from a periodic array is peaked at reciprocal lattice vectors, resulting in a scattering problem that is discrete rather than continuous.  The integrated intensity of each Bragg reflection, then, may be related to a specific Fourier component of the susceptibility of a single, repeating unit, using sum rules familiar from crystallographic structure determination.\cite{warren}  When used in conjunction with resonance techniques \cite{lsco2002} this approach could allow microscopic study of the spatial distribution of electronic states, even in cases where the phase problem cannot be solved.

To enable this approach, we have developed an electron beam lithography technique for patterning large arrays of transition metal oxide nanostructures, which are optimized for RSXS studies.  Illustrated here are macroscopic arrays of 80 nm wide quantum wires of CMR-phase La$_{1-x}$Sr$_x$MnO$_3$ (LSMO) with $x \sim 1/3$, where the LSMO exhibits colossal magnetoresistance (CMR) (Fig. 1).  Comparison to a simple, single-scattering optical model resulted in good, qualitative explanation of the RSXS data.  Some quantitative deviations were, however, observed: A strong background of diffuse scattering, that does not come out of the model, was observed.  We show that this background is due partly to random height variations among the wires, and partly from diffuse magnetic scattering from domains within the wires.  In the best samples, these domains were observed to order, forming a long-ranged pattern with a periodicity of five wires, which likely rises from classical dipole interactions.\cite{puntes2004,chesnel}

\section{Wire Fabrication}

\begin{figure}\centering\rotatebox{0}{\includegraphics[scale=.25]{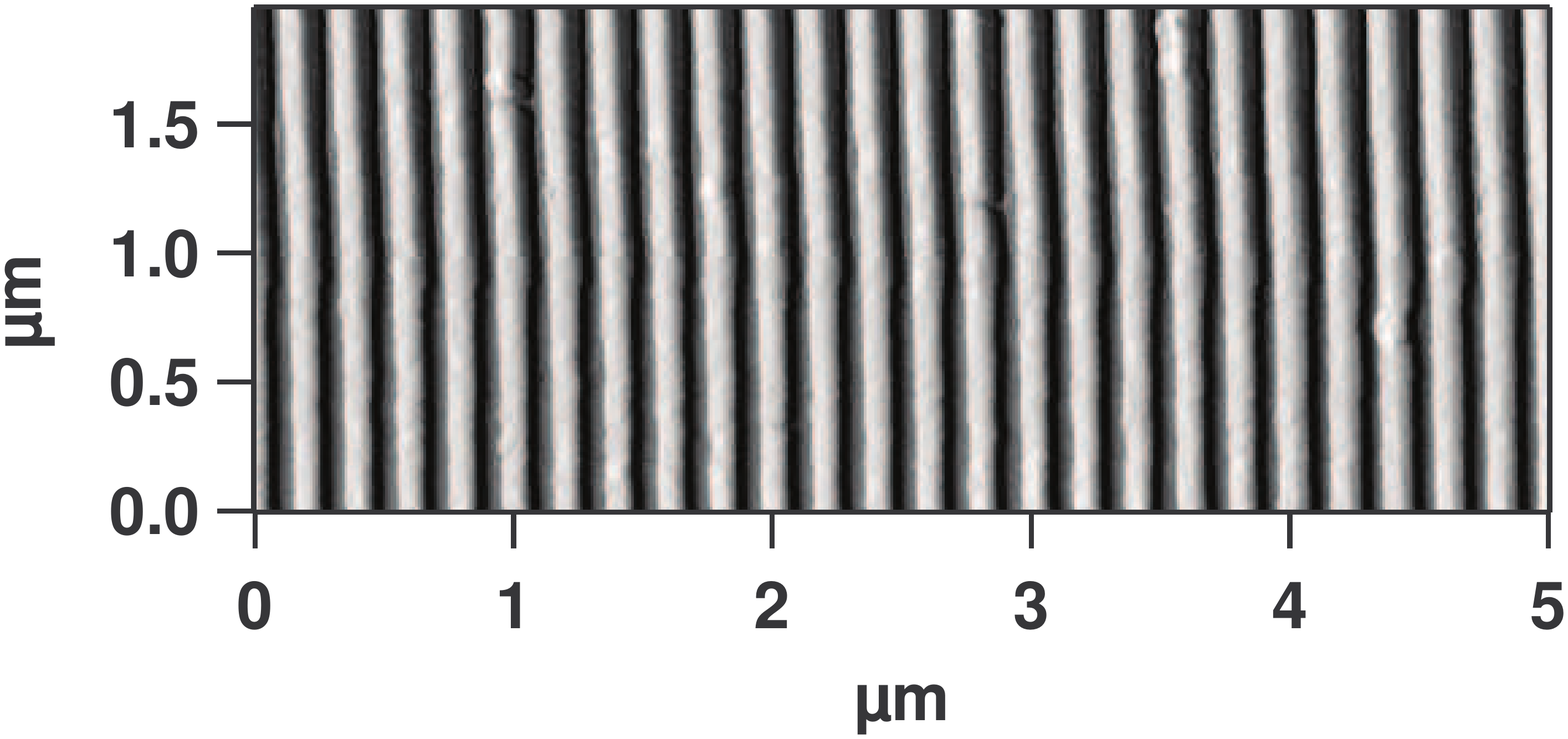}}
\caption{\label{fig:Figure1}  Real space AFM image of sample 2, which gives a periodicity of $d=200$ nm and a wire width $w=80$ nm.  The light regions correspond to the LSMO wires and the dark regions to the spaces between them. 
}
\end{figure}

Arrays of quantum wires were created by a nanofabrication technique employing electron beam lithography followed by ion milling, which is similar to techniques used recently to fabricate arrays of La$_{2-x}$Sr$_x$CuO$_4$ loops.\cite{bozovicLoops}  The starting materials were LSMO alloys and short-period superlattices with average composition $x \sim 1/3$, grown by molecular beam epitaxy (MBE) using techniques described previously.\cite{anand2008}  Three starting structures were used.  First first (``sample 0") consisted of a 25 nm thick LSMO alloy grown on (001)-oriented MgO.  The second (``sample 1") was a digitally ordered superlattice, grown on (001)-oriented SrTiO$_3$ (STO), consisting of 10 periods, each containing two layers of SrMnO$_3$ (SMO) and four layers of LaMnO$_3$ (LMO) (total thickness 21.2 nm).  The last (``sample 2"), also grown on STO, consisted of 27 periods, each containing a single layer of SMO and two layers of LMO (total thickness 31 nm).  The advantage of superlattices, as compared to alloys, is that they exhibit the usual CMR phenomena but with reduced disorder scattering.\cite{anand2008}

To generate wire patterns, the films were spin coated with a layer of PMMA950 A2 resist, which was cured for 10 min at 170 C.  Wire patterns were written, in a series of 100 $\mu$m $\times$ 100 $\mu$m write fields, with a Raith eLine Nano engineering workstation operating in line-exposure mode.  Each pattern consisted of parallel lines, 1 mm in length, spanning a lateral distance of 1 mm.  Slight stitching errors were observed between write fields, but these were too far apart to influence RSXS measurements (see below).  The line pitch was 210 nm in sample 0, 220 nm in sample 1, and 200 nm in sample 2.  The PMMA pattern was developed for 60 sec in a 3:1 solution of MIBK:IPA, rinsed, and dried.  Etch masks 30 nm thick were then deposited using e-beam evaporation.  Al was used for samples 0 and 1, and Ti was used for sample 2.  The expectation was that these would oxidize into Al$_2$O$_3$ and TiO$_2$, respectively, which are hard oxides that would function as good masks during ion milling.  However it was eventually found that even the unoxidized metals function as good masks.  The remaining PMMA was removed by performing acetone lift-off (sonicated for 5 min), leaving patterned masks of Al or Ti metal on top of the LSMO.  

The samples were then placed in an Ar$^+$ ion mill to etch through the LSMO.  Sample 0 was milled for 5 min, which was found to be too long, the result being local patches of wires with most of the surface being bare MgO.  These patches could be studied with local probes such as AFM and magnetic force microscopy (see section IV), but were not suitable for scattering studies.  Samples 1 and 2 were ion milled for 3 min, which was found to etch completely through the LSMO, leaving fully ordered, macroscopic arrays of wires.  Examination with an atomic force microscope (AFM) found the wires in all three samples to have width of approximately 80 nm (see Fig. 1).  

\begin{figure}
\centering\rotatebox{0}{\includegraphics[scale=.3]{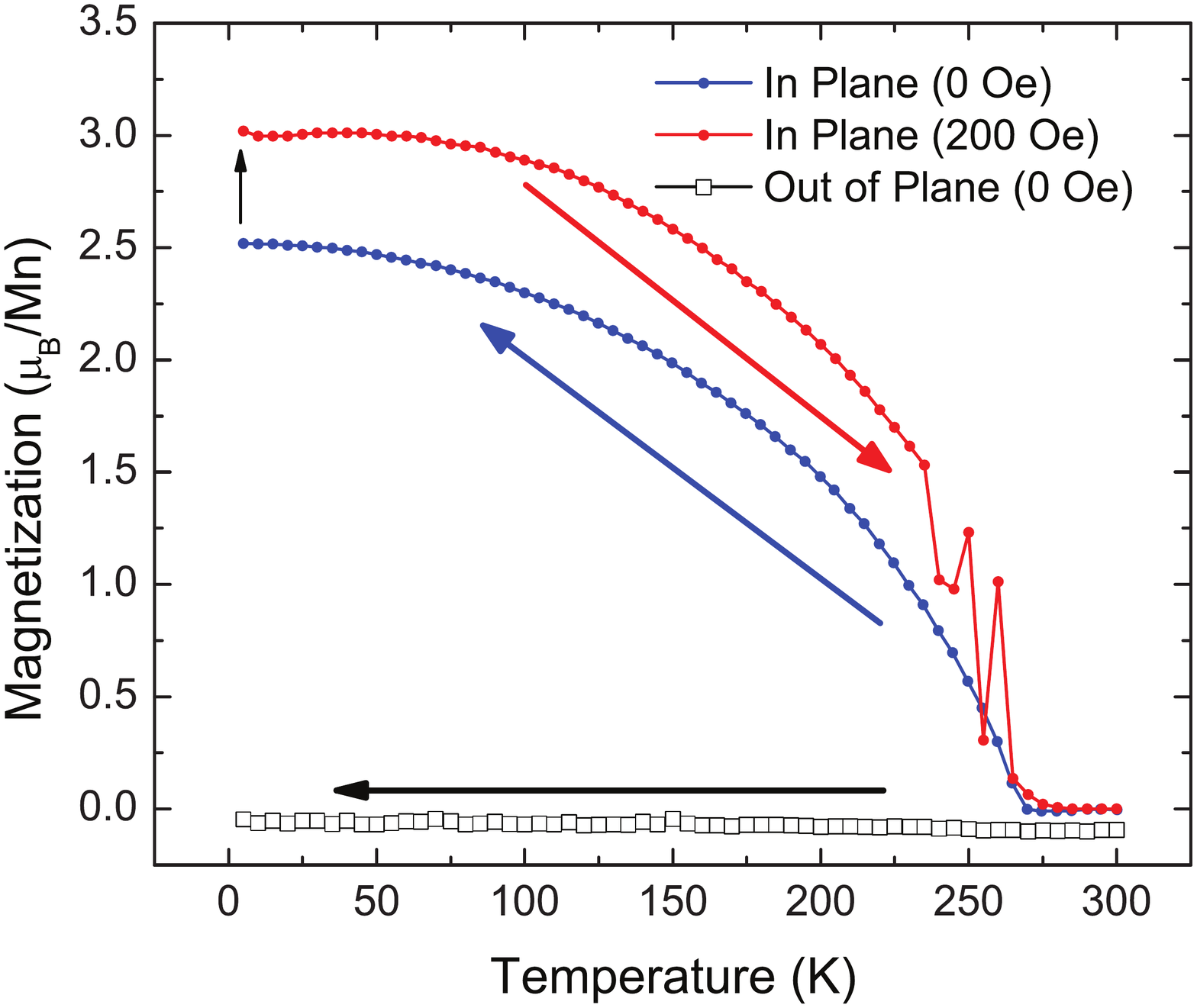}}
\caption{\label{fig:Figure2} Magnetization of sample 1, measured with a SQUID magnetometer (the background from the STO has been subtracted from this measurement).  Measurements of the zero-field-cooled magnetization along the wires (blue line) and perpendicular to the film (black line) revealed the same Curie temperature and anisotropy as unpatterned superlattices of the same composition.  Warming in a 200 Oe field oriented along the quantum wires (red line) exhibited a series of reproducible jumps near $T_C$. 
}
\end{figure}

To check if finite size effects had an obvious influence on the magnetic order, the magnetic properties of sample 2 were characterized with a SQUID magnetometer (Fig. 2).  The Curie temperature and anisotropy, determined with zero-field cooling, were found to be the same as in unpatterned superlattices.\cite{anand2008}  Warming in a 200 Oe field applied along the wires, however, showed reproducible jumps in the magnetization near the Curie temperature that have not been observed in unpatterned films.\cite{anand2008}  Several of the jumps in the field-warmed curve appeared to return to the zero-field cooled measurement, suggesting that there are two distinct magnetic configurations that are very close in energy at this applied field.  The nature of these configurations will be the subject of a forthcoming article.

\section{RSXS Measurements}

\begin{figure}
\centering\rotatebox{0}{\includegraphics[scale=0.3]{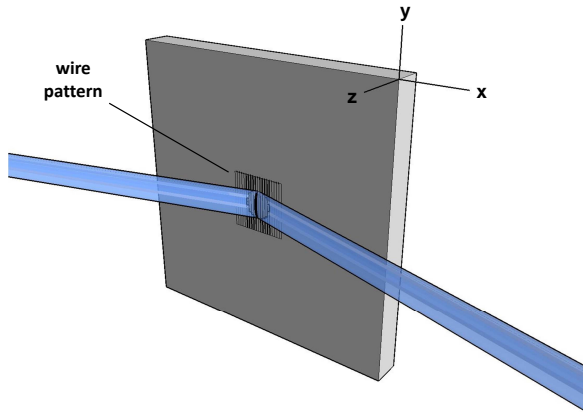}}
\caption{\label{fig:Figure3} Geometry of the RSXS experiments.  The quantum wires are oriented perpendicular to the scattering plane and act like a diffraction grating.
}
\end{figure}

RSXS measurements were carried out at beam line X1B at the National Synchrotron Light Source in an ultrahigh vacuum diffractometer.  Measurements were done with the incident beam tuned near the Mn $L_{2,3}$ edge, and the long axis of the wires oriented perpendicular to the scattering plane (Fig. 3).  We will denote momenta in terms of Miller indices, where $(H,K,L)$ indicates a momentum ${\bf q}=(2\pi H/d,2\pi K/a,2\pi L/t)$.  Here, $d$ is the wire periodicity, $a$ is the lattice parameter of STO, and $t$ is the thickness of the film.  The incident bandwidth was set to 0.2 eV.

\begin{figure}
\centering\rotatebox{0}{\includegraphics[scale=0.4]{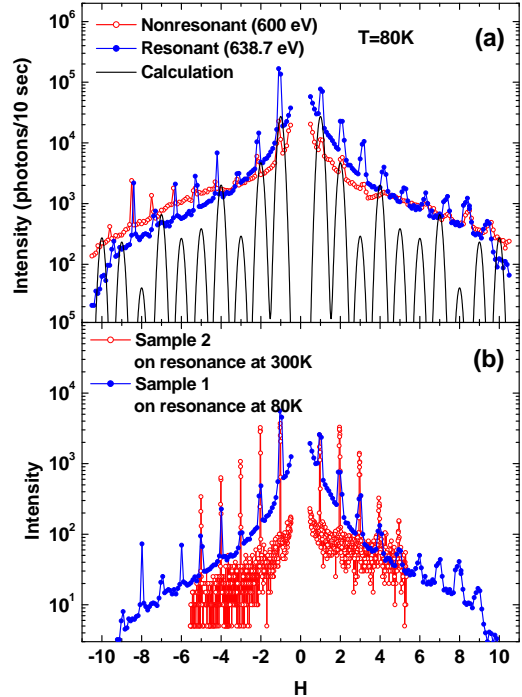}}
\caption{\label{fig:Figure5} RSXS intensity as a function of the in-plane momentum, $H$.  (a) RSXS intensity from sample 1 for $L=7.78$, measured both on (blue points) and off (red points) the Mn $L_3$ resonance (638.7 eV), compared to the analytic model described in section III (black line).  (b) Same resonant data (blue points) compared to that of sample 2 for $L=6.42$ (red line), taken at he Mn $L_2$ edge (649.9 eV). The sharper Bragg peaks and lower diffuse background suggest a higher structural quality of sample 2.
}
\end{figure}

Fig. 4 shows the RSXS intensity for quantum wire samples 1 and 2 as a function of the in-plane momentum, $H$, with the out-of-plane momentum, $L$, held fixed and $K=0$ (scattering measurements were not possible on sample 0).  Measurements were performed both on- and off-resonance, at temperatures both above and below the Curie temperature.  A series of highly resonant peaks was observed at integer values of $H$, which are coherent Bragg reflections from the periodic, wire structure.  While we take a momentum space picture in this paper, these reflections can be thought of as grating reflections, their angular positions defined by the grating equation, 

\begin{equation}
d (\sin{\theta_i} - \sin{\theta_s})=m \lambda,
\end{equation}

\noindent
which is a statement that diffraction maxima occur when the in-plane component of the momentum transfer, $q_x$, is an integer multiple of $2\pi/d$.  We note that these reflections are identical in nature to those observed in a previous study of CoPt quantum wires using ``coherent" x-ray techniques.\cite{chesnel}  In our study these reflections could be observed without the use of coherence filters, since the current geometry (wires $\perp$ to the scattering plane) places the short axis of the momentum resolution ellipse in the $H$ direction.  

\begin{figure}
\centering\rotatebox{0}{\includegraphics[scale=0.25]{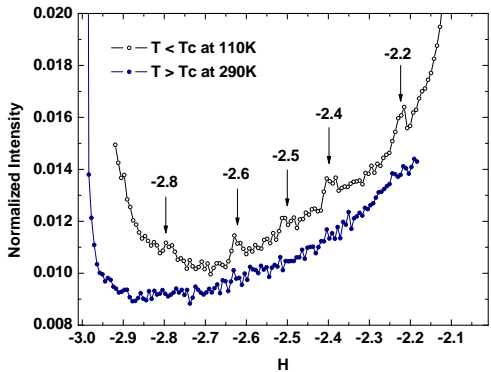}}
\caption{\label{fig:Figure6} RSXS intensity at the Mn $L_2$ edge (649.9 eV) from sample 2, over the range $-3<H<-2$, measured above and below the Curie temperature, $T_C$.  The out-out-of-plane momentum is fixed at $L=6.42$.  Faint superlattice reflections are visible below $T_C$ at the commensurate positions $H=-2.5$ and $H=-3+n/5$, where $n=1,2,3,4$.  These reflections indicate the presence of commensurately modulated antiferromagnetism with a period of five wires.
}
\end{figure}

Close examination of these $H$ scans (Fig. 4) revealed some differences between the two samples, as well as some unexpected observations.  First, the Bragg peaks are much sharper for sample 2 than for sample 1.  Further, both samples exhibited a background of diffuse scattering that decayed rapidly with increasing $|H|$, though this background was higher in sample 1 than in sample 2.  These observations suggest a higher degree of structural perfection of sample 2 compared to sample 1.  We will examine this diffuse scattering in more detail in section IV.

In sample 2, in which the background was weaker, additional superlattice reflections were observed between the Bragg peaks, at the rational fractions $H \pm 1/2$ and $H \pm n/5$, where $H$ and $n$ are integers (Fig. 5).  These reflections were visible only below the Curie temperature, and indicate the formation of a magnetic superstructure with a period of five wires, which we interpret as commensurately modulated antiferromagnetism.  This structure, which most likely arises from classical dipole coupling among the wires,\cite{chesnel} will be the subject of a forthcoming article.

\begin{figure}
\centering\rotatebox{0}{\includegraphics[scale=0.3]{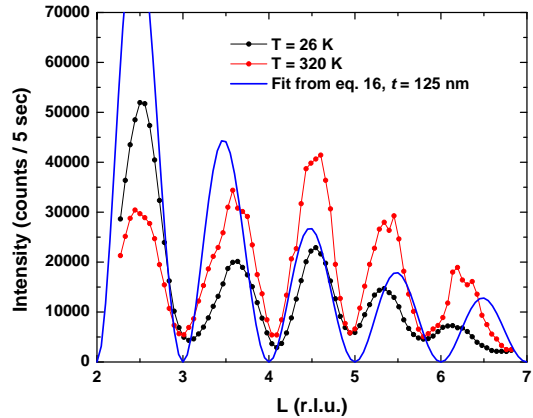}}
\caption{\label{fig:Figure6} RSXS intensity from sample 1, taken resonantly at the Mn $L_3$ edge (638.7 eV), plotted as a function of the out-of-plane momentum, $L$, both above (red points) and below (black points) $T_C$.  In this measurement the in-plane momentum is fixed to $H=1$.  The black line is a fit from the scattering model for $t=21.2$ nm.  Reversible changes in the oscillations are observed as the temperature is cycled through $T_C$.
}
\end{figure}

Fig. 6 shows RSXS measurements of sample 1 as a function of $L$, now with the in-plane momentum fixed at $H=1$.  The scattering along this momentum direction did not exhibit sharp peaks, but oscillated sinusoidally with period $L=1$.  These oscillations are interference fringes that arise from the finite thickness of the structure, and are analogous to the so-called Kiessig fringes characteristic of reflectance from thin, dielectric films.  We emphasize, however, that the measurements in Fig. 6 are not specular; the fringes are part of the out-of-plane momentum structure of the $L=1$ reflection in Fig. 4.  Interestingly, the intensity profile of these fringes, which was measured under resonant conditions, changed reversibly when the temperature is cycled through the Curie temperature of the sample.  We speculate that this is related to the same changing magnetic structure that gives rise to the superlattice peaks in Fig. 5.

\section{Scattering model} 
To account for these features, we introduce a simple, analytically soluble model that provides an intuitive, semi-quantitative understanding of the primary features in the RSXS data.  This model is intended to form the starting point for a forthcoming study of the magnetic structure (i.e., the effect shown Fig. 5).  Here we will make use of the first Born (i.e., single scattering) approximation, in which the scattered intensity is proportional to the square of the Fourier transform of the dielectric susceptibility, $\chi({\bf r})$.\cite{jackson}  We are mainly interested in describing the scattering due to basic charge (as opposed to magnetic) contrast, so we will take $\chi({\bf r})$ to be a scalar.  

Within these assumptions, the periodic array in Fig. 1 can be described by a real space susceptibility

\begin{equation}
\chi=\chi(x,z)
\end{equation}

\noindent
which is a function only of $x$, the coordinate in the plane of the film and perpendicular to the wires, and $z$, the coordinate in the direction normal to the film.  In the experimental geometry in Fig. 3 the momentum transfer has components only in the $x$ and $z$ directions, given by

\begin{equation}
q_x=\left | {\bf q} \right | \sin \left ( \frac{\pi}{2} -\frac{\gamma}{2} +\theta \right )
\end{equation}

\noindent
and 

\begin{equation}
q_z=\left | {\bf q} \right | \cos \left ( \frac{\pi}{2} -\frac{\gamma}{2} +\theta \right ),
\end{equation}

\noindent
where $\gamma$ is the scattering angle, $\theta$ is the sample angle, and the magnitude $|{\bf q}|$ has the value

\begin{equation}
\left | {\bf q} \right | = 2k\sin(\gamma/2).
\end{equation}

Our system is periodic in the $x$ direction but continuous in $z$, so the scattering must be described in terms of a momentum ${\bf q}=(G,q_z)$ where $G=2\pi H/d$ are discrete reciprocal lattice vectors, the Miller index $H$ being an integer, and $q_z$ is a continuous variable describing the momentum perpendicular to the structure.  The general expression for the momentum components of the susceptibility is 

\begin{equation}
\chi^G(q_z)=\frac{1}{d} \int_0^d dx \int_{-\infty}^{\infty} dz \; e^{-i(Gx+q_z z)} \; \chi(x,z).
\end{equation}

\noindent
The quantity $|\chi^G(q_z)|^2$ may interpreted as the $q_z$-dependence of the intensity of the Bragg reflection $G$, integrated along the $q_x$ direction.  

\begin{figure}
\centering\rotatebox{0}{\includegraphics[scale=0.35]{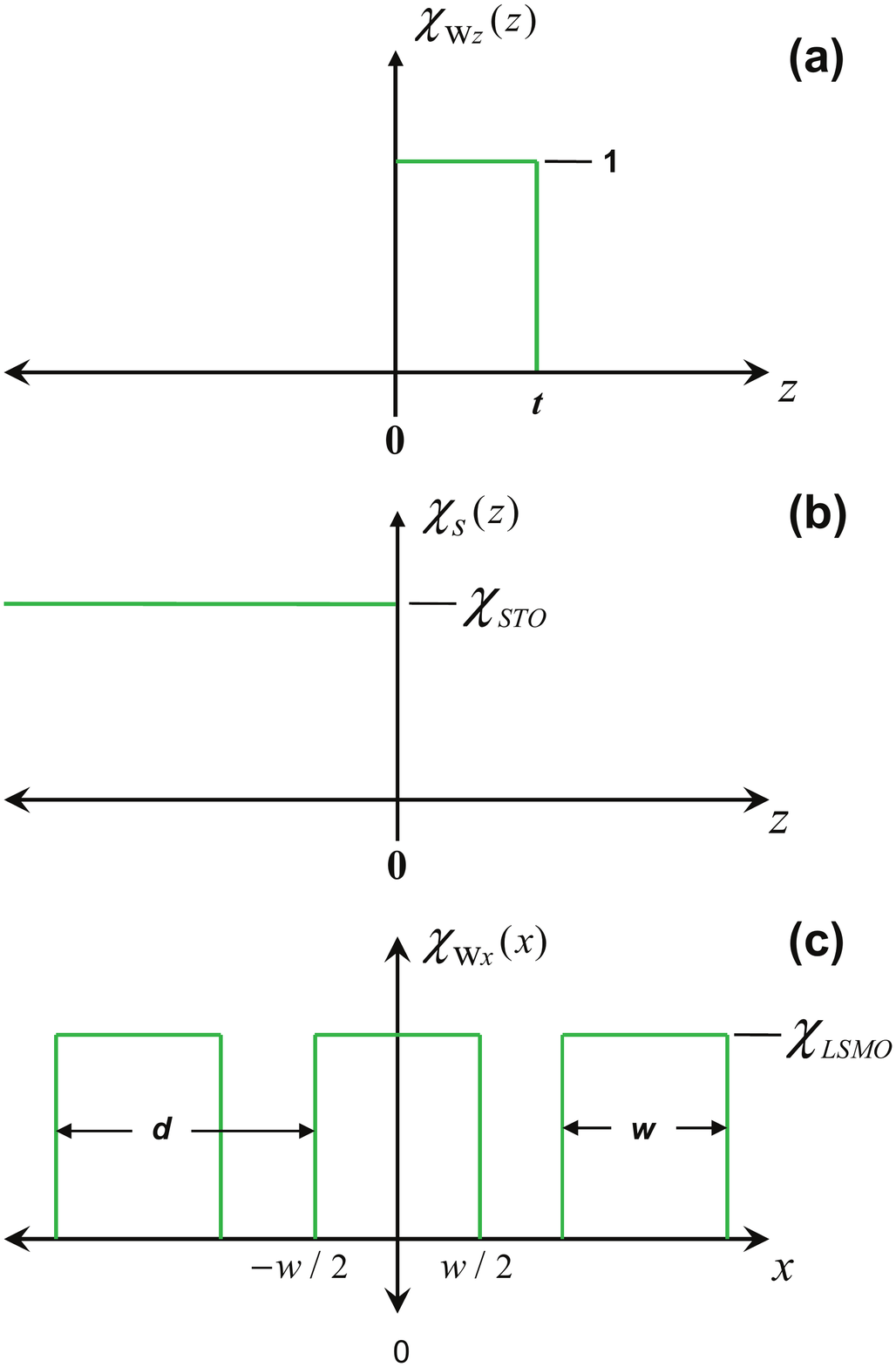}}
\caption{\label{fig:Figure4} Susceptibility functions given in eqs. 9-11.  (a) $\chi_{wz}(z)$ describes the finite thickness of the quantum wires. (b) $\chi_{s}(z)$ is a simple step function, which is the simplest description of a substrate. (c) $\chi_{wx}(x)$ describes the periodic wire structure.  These simple functions, from which the scattered intensity can be computed analytically, capture most of the salient momentum features in the RSXS data.  
}
\end{figure}

Our sample consists of LSMO wires grown on a substrate of STO, and both materials contribute to the
scattering.  Because of the linearity of Born scattering, the problem may be divided up, i.e.,

\begin{equation}
\chi(x,z)=\chi_s(x,z)+\chi_w(x,z)
\end{equation}

\noindent
where $\chi_w$ is the susceptibility of the wires and $\chi_s$ is that of the substrate.

This division allows for some simplifications.  First, the substrate is uniform in the $x$ direction, so 
$\chi_s(x,z)=\chi_s(z)$.  Further, the wire susceptibility is separable, i.e. 

\begin{equation}
\chi_w(x,z)=\chi_{wx}(x)\chi_{wz}(z)
\end{equation}

\noindent
The substrate susceptibility is given by

\begin{equation}
\chi_s(z) = \begin{cases} \chi_{sto} \; e^{\eta z} & z<0 \\ 0 & z>0 \end{cases}
\end{equation}

\noindent
where $\chi_{sto}$ is the uniform susceptibility of STO and $\eta$ is a damping factor that will force convergence in the integrals below.  The sign of the exponent is appropriate for a substrate that fills the half-space $z<0$, which is what we assume here. 

The wires themselves are described by the susceptibility functions

\begin{equation}
\chi_{wz}(z) = \begin{cases} 1  & 0 \leq z\leq t \\ 0 & \mbox{otherwise} \end{cases},
\end{equation}

\noindent
which accounts for the finite thickness of the wires, and 

\begin{equation}
\chi_{wx}(x) = \sum_n{ \begin{cases} \chi_{lsmo}  & -w/2 \leq x+nd \leq w/2 \\ 0 & \mbox{otherwise} \end{cases} },
\end{equation}

\noindent
where $\chi_{lsmo}$ is the uniform susceptibility of LSMO.  This expression accounts for the periodic character of the wire structure.  The various susceptibility functions are plotted in Fig. 7.  

In this framework, the scattering problem can be solved exactly.  The Fourier transforms are readily evaluated analytically:

\begin{equation}
\chi_s^G(q_z)=\frac{\chi_{sto}}{-iq} \, \delta_{G,0}
\end{equation}

\begin{equation}
\chi_{wz}(q_z)=\frac{1-e^{-iqt}}{iq}
\end{equation}

\begin{equation}
\chi_{wx}^G=\chi_{lsmo}\frac{\sin(Gw/2)}{Gd/2}
\end{equation}

\noindent
where $w$ is the wire width and we have taken $\eta \rightarrow 0$.  Piecing the problem back together, the complete structure factor is given by

\begin{equation}
\chi^G(q_z)=\chi_{\mbox{s}}^G(q_z)+\chi_x^G\chi_z(q_z).
\end{equation}

To compare this expression to the experiment we must contend with the fact that, while $G$ is a discrete variable, the in-plane momentum in the experiment, $q_x$, is continuous.  In an actual experiment the discrete Bragg reflections are broadened, by resolution effects, into curves whose intengrated intensity is proportional to $|\chi^G(q_z)|^2$.  Accounting for the effect of finite resolution, the scattered intensity in the experiment is given (to within an overall constant) by 	

\begin{equation}
I(q_x,q_z) = \sum_G \left | \chi^G(q_z) \right |^2 e^{-(q_x-G)^2/\sigma_{q_x}^2}
\end{equation}

\noindent
where the Gaussian function mimics the effect of the momentum resolution of the diffractometer.

In Fig. 4a we compare eq. 16 to the experimental scans along the $H$ (i.e., the $q_x$) direction.  We use the values $d=220$ nm and $w=80$ nm determined with AFM measurements (Fig. 1), and Gaussian width of $\sigma_{q_x}=5 \times 10^{-4} \AA^{-1}$ or, equivalently,  $\sigma_H=0.175$.  Semi-quantitative agreement is attained, eq. 16 producing reasonable values for the positions and integrated intensities of the Bragg reflections, using only rough parameters determined from AFM.

Fig. 6 shows eq. 16 compared to the experimental scans in the $L$ (i.e., the $q_z$) direction, using the value $t=21.2$ nm.  No Gaussian broadening is needed in this direction since $\left | \chi^G(q_z) \right |^2$ is already a continuous function of $q_z$.  Again, reasonable agreement is obtained, the sinusoidal structure of the intensity being reproduced.

An important conclusion that can be drawn from these comparisons is that, except for the features at $H=0$, the momentum-dependence of the RSXS intensity is independent of the homogeneous optical constants $\chi_{lsmo}$ and $\chi_{sto}$.  These quantities enter only as prefactors that affect the overall intensity, there being no explicit interference between them.  While this model, as it is currently implemented, only accounts for the course charge features in the data, i.e., the locations of the Bragg reflections and the form of the oscillations etc., it may easily be generalized to account for magnetic scattering by using an appropriate tensorial form for the susceptibility, $\chi(x,z)$.  This approach will be used to analyze the more exotic features, such the magnetic superlattice shown Fig. 5.  

\section{Diffuse Scattering} 
In this section we address the origin of the diffuse scattering, which is most clearly visible in Fig. 4.  This background is stronger in sample 1, which was fabricated using an Al etch mask, than in sample 2, which was masked with Ti.  Further, the Bragg reflections from sample 1 are weaker and broader than in sample 2.  These observations suggest that the diffuse scattering arises, at least partly, from imperfections in the wire pattern itself.

\begin{figure}
\centering\rotatebox{0}{\includegraphics[scale=0.3]{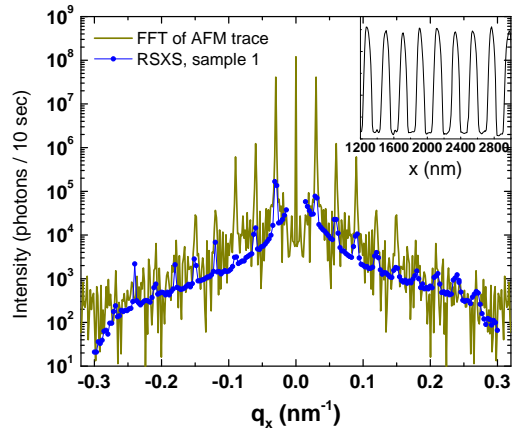}}
\caption{\label{fig:Figure7} Comparison of the RSXS data from sample 1 (also shown in Fig. 4) to the square of a Fast Fourier Transform (FFT) of AFM data taken from a $5 \mu m \times 5 \mu m$ section of the same sample.  The same diffuse background is observed in both measurements.  (inset) Real space AFM trace showing random height variations among the wires, which is the origin of the diffuse signal.
}

\end{figure}

To evaluate this possibility, we performed AFM measurements of sample 1 over a large ($5 \mu$m $\times$ $5 \mu$m) field of view, which could then be Fourier transformed for comparison to scattering experiments.  Close inspection (Fig. 8, inset) indeed revealed a source of disorder: random height variations among the wires, most likely a result of nonuniformity in deposition of the masks or in ion milling, were observed.

To determine if these variations were the source of diffuse scattering, the AFM image was Fourier transformed, squared, and compared to the RSXS data from sample 1 (Fig. 8).  As expected, sharp peaks due to the periodic wire structure are observed at the same $q_x$ values as in the RSXS experiments.  Importantly, diffuse ``scattering" is also observed, due to the random wire heights, which has the same dependence on $q_x$ as the RSXS data.  This suggests that the diffuse background is a result of nonuniformity in either the deposition of the mask or the ion milling, and is more severe when using Al, rather than Ti, as an etch mask.  This disorder is strong enough to wash out the interesting magnetic correlations observed in sample 2 (Fig. 5).

\begin{figure}
\centering\rotatebox{0}{\includegraphics[scale=0.35]{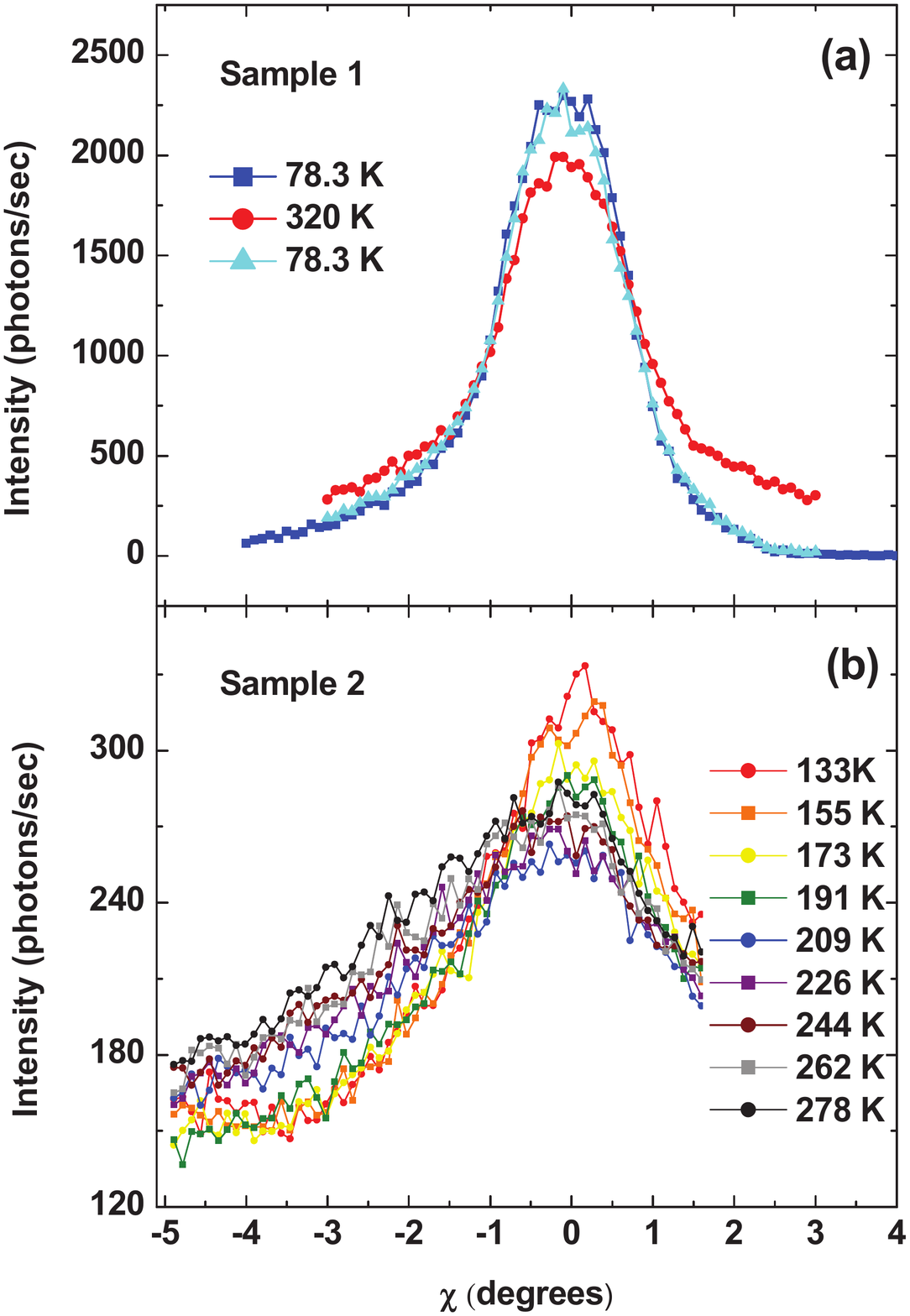}}
\caption{\label{fig:Figure8} Temperature dependence of diffuse RSXS scattering from both samples, measured along the $K$ momentum direction, by scanning the sample $\chi$ angle (i.e., rotating the sample around the $x$ axis defined in Fig. 3).  (a) Data from sample 1, taken at the Mn $L_3$ edge at fixed $H=0.44$ and $L=4.88$, showing reversible changes in the width of the diffuse scattering.  (b) The same measurement for sample 2, taken at the Mn $L_2$ edge at fixed $H=-2.5$ and $L=6.42$, which also shows reversible width changes.  These changes are interpreted as arising from magnetic domain formation within the wires with increasing temperature.  The changes in sample 1 are less pronounced because a higher proportion of the diffuse scattering is from imperfections in the structure.
}
\end{figure}

Not all the diffuse scattering is from disorder, however.  In Fig. 9 we show scans through the diffuse scattering in the $q_y$ direction (i.e., the $K$ direction), which were accomplished by placing $H$ between two Bragg reflections and scanning the $\chi$ sample motion.  The diffuse scattering was found to peak strongly at $\chi=0$ or, equivalently, $K=0$, its width$-$to first order$-$being a measure of the height correlation length along the wires.  Surprisingly, we found this width to be slightly temperature-dependent in both samples 1 and 2 (Fig. 9).  As the temperature was cycled through the Curie temperature, the width of the diffuse scattering changed by approximately 10 \% in sample 1 and 40 \% in sample 2. Because the height variations cannot depend on temperature, we interpret this result as evidence that, while most of the diffuse scattering comes from structural disorder (particularly in sample 1), some amount of the scattering arises from random magnetic domains in the wires.  The size of these domains changes as the temperature is cycled through $T_C$.  These changes are likely related to the effects observed in Figs. 5 and 6.

\begin{figure}
\centering\rotatebox{0}{\includegraphics[scale=0.28]{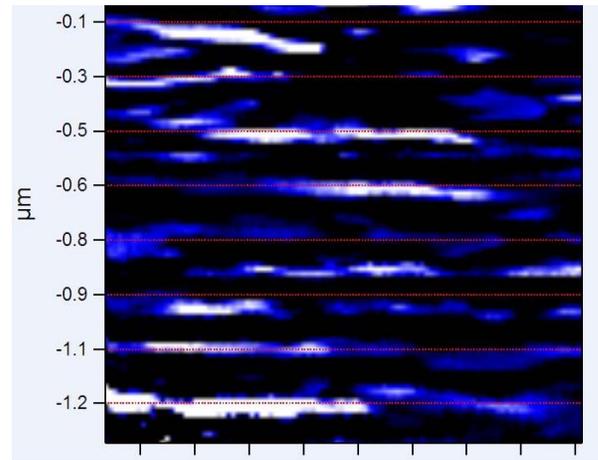}}
\caption{\label{fig:MFM} False color plot showing the cantilever frequency shift, with the white regions corresponding to larger frequency shift.  The color range, spanning black to white, corresponds to a shift in the cantilever frequency of 6 Hz. The cantilever used for this study was a MikroMasch model CSC12 tipless AFM probe with stiffness $k = 0.03$ N/m and frequency $f = 10$ kHz. The red solid lines, separated by 200 nm [SHOULD BE 210], are placed as a guide to the eye to indicate the pitch of the LSMO wires.
}
\end{figure}

To investigate the possible existence of magnetic domains, we performed a magnetic force microscopy (MFM) measurement on one of the local patches of wires in sample 0 (Fig. 10) at $T = 4 K$.  For the MFM measurements, a micron-sized samarium-cobalt magnet was glued to the end of a low stiffness, Si$_3$N$_4$ cantilever, with the magnetic moment of the particle aligned perpendicular to the cantilever axis. The particle was shaped using a focused ion beam to a tip diameter of 100 nm. To image the in-plane component of the magnetic moment, the cantilever axis was oriented perpendicular to the sample surface, and the frequency shift of the cantilever was measured. For a tip-surface separation of 100 nm, it was found that the frequency shift, indicated as the false color image (Fig. 10), is dominated by the magnetic tip-sample interactions and not the sample topography. Magnetic domains, of order 200 nm in size, were clearly observed along the wires. Such domains should give rise to diffuse magnetic scattering, and could be the origin of the width changes in Fig. 9. Because these measurements were done at fixed temperature it was not possible to directly confirm that the changes were a result of changing domain size. But these results clearly establish that these structures exhibit a non-trivial domain structure, even at the lowest temperatures measured. The details of the magnetic structure will be the subject of a forthcoming article.

\section{summary}
In summary, we have developed a fabrication procedure for creating very large arrays of transition metal oxide quantum wires$-$in this case of CMR-phase La$_{1-x}$Sr$_x$MnO$_3$$-$that are suitable for resonant scattering experiments.  The strategy of creating many copies of the structure of interest, by analogy with x-ray crystallography techniques, results in a simplified scattering problem in which the integrated intensity the in-plane Bragg reflections may be related to individual Fourier components of the susceptibility of an (average) isolated structure.  Use of a Ti mask during ion milling was found to result in higher quality patterns, as defined by the amount of diffuse disorder scattering, than use of an Al mask.  An exactly soluble, analytic model of the scattering was introduced that captures, semi-quantitatively, the primary features in the RSXS data, providing a simple framework for analyzing future measurements of this type.  Resonant magnetic x-ray scattering experiments, carried out at the Mn $L_{2,3}$ edge, indicated a highly non-trivial pattern of magnetic domains that, in sample 2, were observed to order with a commensurate period of five wires.  The details of this magnetic order will be described in a forthcoming article.

We acknowledge Lane Martin for helpful discussions.  This work was supported by the U. S. Department of Energy, Office of Basic Energy Sciences. 
Wire fabrication was supported by contract DE-FG02-07ER46453 through the Frederick Seitz Materials Research Laboratory.  RSXS studies were supported by contract no. DE-FG02-06ER46285, with use of the NSLS supported under grant no. DE-AC02-98CH10886.  Superlattice growth at the Center for Nanoscale Materials was supported by contract no. DE-AC02-06CH11357.

\bibliography{wires}

\end{document}